\documentclass[]{aa}

\usepackage{graphicx}             
\usepackage{float}                 
\usepackage{placeins}              
\usepackage{adjustbox}             
\usepackage{lscape}                
\usepackage{wrapfig}               
\usepackage{epstopdf}              
\usepackage{subfigure}             
\usepackage{booktabs}              
\usepackage{xcolor}                
\usepackage{amsmath}               
\usepackage[T1]{fontenc}           
\usepackage{float}                 
\usepackage{gensymb}               
\usepackage{rotating}              
\usepackage{siunitx}
\usepackage{dcolumn}
\usepackage{hyperref}

\usepackage{xcolor} 
\usepackage{txfonts}
\usepackage{natbib}
\usepackage{graphicx}  
\usepackage{amsmath}   
\usepackage{multirow}  

\bibpunct{(}{)}{;}{a}{}{,} 

\DeclareRobustCommand{\VAN}[3]{#2}
\let\VANthebibliography\thebibliography
\def\thebibliography{\DeclareRobustCommand{\VAN}[3]{##3}\VANthebibliography}

\DeclareGraphicsExtensions{.ps}
\usepackage{graphicx}
\usepackage{txfonts}
\begin{document}

\title{A thousand looks at X Persei: X-ray spectroscopy at high time resolution}

\subtitle{}

\author{
Sanjurjo-Ferr\'{i}n, G.$^{1}$,
Torrej\'on, J.M.$^{1}$,
Rodes-Roca, J.J.$^{1}$,
Postnov, K.$^{2}$,
Planelles-Villalva,J.$^{1}$,
Oskinova, L.$^{3}$,
}

\institute{
$^{1}$Instituto Universitario de F\'{i}sica Aplicada a las Ciencias y las Tecnolog\'ias, Universidad de Alicante, 03690 Alicante, Spain\\
$^{2}$ Sternberg Astronomical Institute, Moscow M.V. Lomonosov State University, Universitetskij pr, 13, Moscow 119234, Russia\\
$^{3}$Institute for Physics and Astronomy, University of Potsdam, 14476 Potsdam,
Germany}

   \date{}

\abstract
{X Persei is a classical Be/X-ray binary composed of an O9.5III-B0V Be type star and a neutron star (NS). The NS exhibits coherent pulsations with a spin period of $\sim 837\, \mathrm{s}$, orbiting the Be star with $\sim 250\, \mathrm{d}$ period. X Persei is notable for its exceptionally hard X-ray emission extending beyond 100 keV. Due to the mild eccentricity of the orbit $\sim$ 0.11, the orbital separation varies between roughly 35 R${_\star}$ (periastron) and 44 R${_\star}$ (apastron).}
{In this work, we analyze five targeted observations obtained with the \textit{XMM-Newton} and \textit{Chandra} observatories taken in a 10-year time span, with the aim of investigating the structure and variability of the circumstellar disk surrounding the Be star, in particular the presence of over-dense areas known as clumps.}
{We performed spectral and timing analyses, including average and NS spin resolved spectroscopy for three of the observations, producing individual spectra at intervals as short as 210 s, corresponding to different epochs of the NS spin, resulting in approximately 1200 spectra. 
This detailed analysis aimed at resolving emission line features otherwise diluted in averaged spectra and the evolution of continuum components along the NS spin. 
}
{The observed spectra are accurately modeled by a two-component continuum comprising a high-temperature blackbody and a powerlaw component.  Phase resolved spectroscopy reveals transient Fe K$\alpha$ emission linked to clumps in the circumstellar disk during high density epochs, with a 8-9\% prevalence. Dips in the X-ray lightcurve are tied to clump passages. The disk density modeling, based solely on X-ray data, suggests a compact and dense disk, with a radial density exponent $\alpha$ of 2.4–3.3, and a high inner disk density, $\rho_{0}\sim(6$–$20)\times10^{-10}\,\mathrm{g\,cm^{-3}}$, in agreement with previous studies conducted in the optical and infrared bands.}{}


\titlerunning{X Persei}

\authorrunning{Sanjurjo-Ferr\'{\i}n et al. }

\maketitle

\section{Introduction}

In Be X-ray binaries (BeXB), the optical companion is a Be star. Classical Be stars are rapidly rotating, near–main‑sequence B‑type stars of luminosity class III–V whose optical spectra show prominent Balmer (and often metallic) emission lines, hence the 'e' suffix in their spectral designation. These emission features arise from excess radiation produced by a gaseous circumstellar environment, generally interpreted as a geometrically thin, quasi‑Keplerian disk in the stellar equatorial plane \citep{2003PASP..115.1153P}.

On average, the disks surrounding Be stars in binary systems are approximately twice as dense as those around isolated Be stars. This increased density results primarily from tidal truncation caused by gravitational interaction with the compact object, which reduces the size of the circumstellar disk \citep{2001A&A...367..884Z}. Consequently, BeXB systems exhibit smaller yet denser equatorial disks compared to their isolated counterparts \citep{2011Ap&SS.332....1R}. \cite{2025A&A...698A.309R} demonstrated with smoothed particle hydrodynamics simulations that circumstellar matter can be divided into five regions of interest: an inner Be disk, spiral dominated disk, and bridge, as well as circumsecondary and circumbinary regions. Thus, although the circumstellar envelope is traditionally considered to be smooth, it actually has structure. Furthermore, it may exhibit the small-scale overdensities present in the stellar winds of massive stars, known as wind clumps. Investigating this possibility is one of the goals of the present paper. The presence of clumps in stellar winds of early type stars had been known and studied for a long time. Instabilities in their line-driven winds lead to the formation of shocks and inhomogeneities—commonly referred to as clumps \citep{1984ApJ...284..337O,1988ApJ...335..914O}. 
\cite{2020A&A...634A..49H} derived clumping factors for a sample of High Mass X-ray Binaries (HMXBs), showing similar characteristics as the isolated massive stars. 

The X‑ray source \object{4U~0352+309} was first detected by the \textit{Uhuru} satellite \citep{1972ApJ...178..281G} and was later identified as the companion of the BeO9.5 star X~Persei \citep{1972Natur.235..273B}.  Timing analyses reveal coherent X‑ray pulsations with a period of $\sim\!837\ \mathrm{s}$ \citep{1976MNRAS.176..201W,1998ApJ...509..897D}, demonstrating that the accretor is a neutron star (NS) \citep{Delgado-Martí_2001}.  The binary orbit
is slightly eccentric ($e\approx0.11$) with a period of $\sim\!250\ \mathrm{d}$ and an inclination of $i\simeq26^{\circ}\!-\!33^{\circ}$ \citep{Delgado-Martí_2001}.  Estimates of the distance from \textit{Gaia}~DR3  \citep{2023A&A...674A...1G} yields a parallax-based distance of $0.690\pm 0.011$ kpc.  Note that the re-normalized unit weight error for this value is 1.5, when the acceptance value proposed by Gaia is $<$ 1.4.

\textit{INTEGRAL} data confirm that X Persei is exceptionally hard, remaining visible beyond 100 keV \citep{2012A&A...540L...1D}. Its 4–200 keV spectrum separates cleanly into two components that are well described by thermal + bulk Comptonization models near the NS surface. Consistent with other BeXB, \cite{2024AA689A186R} observed a 'two hump' spectral shape during a low-accretion state in X Persei,  showing a pronounced high‑energy tail that cannot be explained by Comptonization. Its broad $\sim$ 27 keV cyclotron‑resonant scattering feature likewise rules out a Cyclotron Resonant Scattering Feature (CRSF) origin, pointing instead to direct cyclotron emission as the most plausible mechanism for the hard spectral component. Complementary torque analysis of long-term spin evolution favors a magnetar-strength field of order 10$^{14}$ G \citep{2012A&A...540L...1D}.
Consistently, \cite{2018PASJ...70...89Y} using the Ghosh \& Lamb model \citep{1979ApJ...234..296G} suggested that the magnetic field should be in the $(4-25)\times $ 10$^{13}$ G range and that the NS mass is (2.03 $\pm$ 0.17) $M_{\odot}$. In this paper, we will instead use the NS mass 1.6 $M_{\odot}$ derived in \cite{Delgado-Martí_2001}.

The results with the \textit{Imaging X-ray Polarimetry Explorer} show that the X-ray polarization signal is strongly dependent on the spin phase of the pulsar and the large angle between the rotation and magnetic dipole axes \citep{2023MNRAS.524.2004M}. \citet{2020MNRAS.499.3650Z} reported an eccentric density wave in the circumstellar disk of \object{4U~0352+309}.  
This wave propagates outward at a velocity of \(v_{\text{wave}} = (1.1 \pm 0.2)\,\mathrm{km\,s^{-1}}\) and has an eccentricity in the range \(0.17 \le e_{\text{wave}} \le 0.41\).

\begin{table}
\centering
\caption{Main astrophysical parameters of X~Persei.}
\label{parameters}
\begin{tabular}{l c}
\hline
\multicolumn{2}{c}{\textbf{Companion}}\\
\hline
Spectral type & O9.5III \tablefootmark{a}\\
Radius ($R_{\odot}$) & 5--10\tablefootmark{b}\\
Mass ($M_{\odot}$) & 13--20\tablefootmark{b}\\
\hline
\multicolumn{2}{c}{\textbf{Neutron star}}\\
\hline
Mass ($M_{\odot}$) & 1.6\tablefootmark{c}\\
\hline
\multicolumn{2}{c}{\textbf{Binary system}}\\
\hline
$d$ (kpc) & 0.69\tablefootmark{d}\\
$L_{X}$ ($\times 10^{36}\,\mathrm{erg\,s^{-1}}$) & 0.1--1.3 (this work)\\
$P_{\rm orb}$ (days) & $250.5\pm0.6$\tablefootmark{c}\\
$P_{\rm pulse}$ (s) & $837.6713\pm0.0003$\tablefootmark{c}\\
$E(B-V)$ & 0.39\tablefootmark{e}\\
Inclination ($^\circ$) & $\sim$26--33\tablefootmark{c}\\
Eccentricity & $0.111\pm0.018$\tablefootmark{c}\\
Angle to periapsis ($^\circ$) & $108\pm9$\tablefootmark{c}\\
\hline
\multicolumn{2}{c}{\textbf{Circumstellar disk}}\\
\hline
$\alpha$ &  2.4--3.3 (this work) \\
$\log_{10}\rho_{0}$ (g\,cm$^{-3}$) & $-(8.7$--$9.2)$(this work)\\
\hline
\end{tabular}

\tablefoot{
The $\alpha$ and $\rho_{0}$ ranges are obtained for the dataset; they represent the density exponent and the disk density close to the stellar surface for a viscous decretion disk model \citep{2017A&A...601A..74K}. See Sec.~\ref{sec:disc} and Table~\ref{table:disk_parameters_ps} for individual values.
\tablefoottext{a}{\citet{1982ApJS...50...55S}.}
\tablefoottext{b}{\citet{1997MNRAS.286..549L,1992MNRAS.258..439R}.}
\tablefoottext{c}{\citet{Delgado-Martí_2001}.}
\tablefoottext{d}{\citet{2023A&A...674A...1G}.}
\tablefoottext{e}{\citet{1998MNRAS.296..785T}.}
}
\end{table}

 This paper is organized as follows. First, we describe the observations analyzed and the software used. In the Results section, we present the average spectral analysis, NS spin properties, and the NS spin–resolved analysis, which comprises approximately 1200 spectra. We then address the timing analysis of features identified within the light curves, such as clumps and spikes. The Discussion section is divided into two main parts, the first focuses on the structure of the Be stellar disk, while the second examines the NS emission.

\section{Observations and analysis}

\begin{table*}
\centering
\caption {Observation log.}

\begin{tabular}{ccccccc}
\hline
 & Obs id   & Instrument & Grating&  Date         & Duration(ks) & Orbital phase \\\\
 \hline
Chandra-1  & 589       & HRC-S &LETG& 2000-09-27 &  50 & 0.94\\
XMM-1      & 0151380101 & PN  &  -  & 2003-02-25 &  32 & 0.41\\
Chandra-2  & 8229      & ACIS-I&  - & 2007-10-28 &  50 & 0.25\\
XMM-2      & 0600980101 & PN   & -    & 2010-02-23 & 126 & 0.60\\
Chandra-3  & 12447     & ACIS-S & LETG & 2010-12-20 & 124 & 0.83\\
\hline
\\
\end{tabular}%

\label{table:obs-log}
\end{table*}

In this paper, we analyze five different observations, performed with two different X-ray observatories, \textit{XMM-Newton} and \textit{Chandra}. The \textit{Chandra} telescope, launched by NASA in 1999, employs a system of four nested grazing-incidence mirrors coated with iridium. These mirrors focus X-rays onto detectors with an angular resolution of 0.5 arcseconds. \textit{Chandra} operates over an energy range of ($0.1$–$10$) keV using two primary instruments: the High Resolution Camera (HRC) for imaging and the Advanced CCD Imaging Spectrometer (ACIS) for detailed spectroscopy \citep{Weisskopf_2000}. 

The \textit{XMM-Newton} spacecraft, launched by ESA in 1999, carries a set of three X-ray CCD cameras, known as the European Photon Imaging Camera (EPIC). Two of the cameras are Metal Oxide Semi-conductor (MOS) CCD arrays. They are installed behind the X-ray telescopes that are equipped with the gratings of the Reflection Grating Spectrometers (RGS).The third X-ray telescope uses pn CCDs and is referred to as the pn camera. The energy range goes from 0.15 to 15 keV \citep{2001A&A...365L..27T,2001A&A...365L..18S}.

The \textit{XMM-Newton} observations were performed with the European Photon Imaging Camera (EPIC) pn detector, which provides a high effective area in the (0.3--10)~keV energy range and moderate spectral resolution ($E/\Delta E \sim 20$--50). 
The \textit{Chandra} observations were obtained with three different configurations: the High Resolution Camera Spectroscopy detector with the Low Energy Transmission Grating (HRC-S/LETG), the Advanced CCD Imaging Spectrometer imaging array (ACIS-I), and the Advanced CCD Imaging Spectrometer spectroscopy array with the Low Energy Transmission Grating (ACIS-S/LETG). The LETG grating observations offer high spectral resolution ($E/\Delta E \gtrsim 500$ at 1~keV). In contrast, ACIS-I provides moderate spectral resolution ($E/\Delta E \sim 7$--10 at 1~keV) and limited sensitivity.

\begin{figure}
\centering
\subfigure{\includegraphics[trim={0cm 0cm 0cm 0cm},width=1.0\columnwidth]{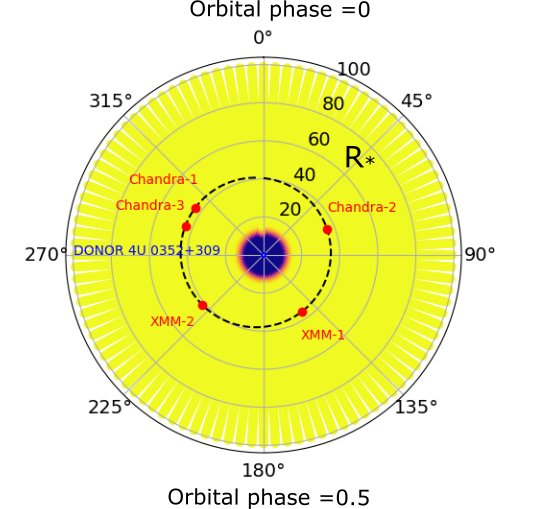}}
\caption{Pole-on sketch of the system showing the relative sizes of the companion star and the NS orbit. Disk density is color-coded, with darker shades representing higher density. The observations are marked on the orbit at their respective orbital phases,  where 180$^\circ$ corresponds to orbital phase 0.5 and 0$^\circ$ orbital phase 0. Ephemeris were taken from \cite{Delgado-Martí_2001} and 180$^{\circ}$ represents orbital phase 0.5 (towards the observer). }
\label{fig:sketch}
\end{figure}

The observation log is collected in Table \ref{table:obs-log}. A sketch of the system, the orbit and the orbital phase of each observation is displayed in Fig. \ref{fig:sketch}. The color gradient represents the density of the disk, with darker shades representing the higher density.

For the \textit{Chandra} data we used standard \textsc{CIAO} procedures (v4.17) with CALDB 4.11.1 \citep{2006SPIE.6270E..1VF}. We reprocessed the data and extracted spectra following the \textsc{CIAO} threads. For the grating observations (Chandra-1: HRC-S/LETG and Chandra-3: ACIS-S/LETG), we extracted the first-order LETG spectra ($m=\pm1$), built the corresponding response files (ARF and RMF), and combined the $+1$ and $-1$ orders to increase the signal-to-noise ratio. For the imaging observation (Chandra-2: ACIS-I), we extracted source and background spectra and generated the appropriate responses using the standard pipeline.

For the \textit{XMM-Newton} observations, the data was reduced with the \textsc{sas} software v22.1 \citep{2017xru..conf...84G}. The task \texttt{epatplot} revealed that the three EPIC cameras were affected by pile-up. To mitigate this effect, we excluded the most affected central pixels from the source region. After correction, only the pn data met the required quality criteria; consequently, the present analysis is based solely on the pn camera. We kept events with \texttt{flag = 0} and a pattern between 0 and 4. Data were filtered through \texttt{\#XMMEA EP} \citep{2001A&A...365L..27T}.

The spectra were analyzed and modeled with the  Interactive Spectral Interpretation System package (\textsc{isis})\footnote{ \url{https://space.mit.edu/cxc/isis/}}. The emission lines were identified using the \textsc{atomdb} data base package\footnote{\url{http://www.atomdb.org/}} and \texttt{XSTAR} \citep{2021Atoms...9...12M}. 

Orbital modulations in this data set were analyzed with the Python package \texttt{xraybinaryorbit} \citep{Sanjurjo-Ferrin2024}\footnote{\url{https://xragua.github.io/xraybinaryorbit/}}. Emission lines were identified with our publicly available \texttt{ISIS}-based tool \texttt{BLISS}\footnote{\url{https://xragua.github.io/BLiSS/}}. 
The main feature of this automatic line-detection pipeline is that it is designed to be  continuum-independent, and can therefore be applied to spectra of arbitrary shape. Within \texttt{ISIS}, the code iterates over the detected candidate lines and retains those that produce an improvement in the reduced $\chi ^{2}$ above the chosen threshold. To detect relevant features in the light curves, such as dips and spikes, we used the Python package \texttt{dipspeaks} \citep{dipspeaks_zenodo} publicly available online\footnote{\url{https://github.com/xragua/dipspeaks}}. The method generates synthetic light curves that reproduce the noise properties of the data to characterize the expected distribution of non-astrophysical fluctuations; significant deviations from this behavior in the real light curves are then flagged as genuine astrophysical events.

\section{Results}

 The observations analyzed in this study are acquired with different observatories, instrumental setups, and sensitivities. The analysis methods are tailored to each dataset. For clarity, Table~\ref{tab:obs_usage} provides an overview of the observations used in each part of the analysis and, where applicable, the primary reasons for their exclusion.

\begin{table*}
\caption{  Summary of the observations used in each analysis step and the main reason for exclusion when applicable.}
\label{tab:obs_usage}
\centering
\small
\begin{tabular}{lcccccc}
\hline\hline
Observation & Average spectral & Spin period / & NS spin resolved & Fe K$\alpha$ search & Dips & Spikes \\
            & analysis         & folded pulse  & continuum      & in 210 s spectra    &      &        \\
\hline
Chandra-1 & Used & Used$^{a}$  & Used        & Not used$^{b}$ & Not used$^{d}$ & Not used$^{e}$ \\
XMM-1     & Used & Used        & Used        & Used           & Used           & Used \\
Chandra-2 & Used & Used        & Used          & Not used$^{c}$ & Used & Used \\
XMM-2     & Used & Used        & Used        & Used           & Used           & Used \\
Chandra-3 & Used & Used        & Used        & Used           & Used           & Used \\
\hline
\end{tabular}
\tablefoot{ \tablefoottext{a}{Only the soft-band folded pulse is shown, since the HRC-S/LETG configuration has limited sensitivity at higher energies and no intrinsic spectral resolution.}
\tablefoottext{b}{Excluded because HRC-S/LETG lacks intrinsic energy resolution and its effective area decreases strongly toward Fe K energies.}
\tablefoottext{c}{Line search was attempted, but the spectral resolution and signal-to-noise ratio of the individual 210 s spectra were insufficient to detect emission lines.}
\tablefoottext{d}{Not used in the dip analysis because the adopted dip definition requires simultaneous soft (0.3-3 keV) and hard (3-10 keV) light curve coverage.}
\tablefoottext{e}{Not used in the spike analysis, as this analysis require hard band (3-10 keV) light curve coverage.}}

\end{table*}

\subsection{Spectral analysis}

To perform the spectral analysis we re-binned all the spectra analyzed through this paper to achieve at least 25 counts per bin, ensuring that $\chi^{2}$ statistics are valid while preserving sufficient energy resolution for line searches.
We fitted each of the five observations with an absorbed \texttt{powerlaw} plus a \texttt{blackbody} component. 
A very similar model was applied by \cite{2007A&A...474..137L} to the XMM-1 observation, yielding comparable parameters.
The parameters of the \texttt{bbody} model include the temperature ($kT_{\rm bb}$) and normalization, defined as $L_{39} D_{10}^{-2}$ where $D_{10}$ is the distance to the source in units of 10 kpc and $L_{39}$ is the luminosity in units of $10^{39}\,\mathrm{erg\,s^{-1}}$.

The absorption was modeled using the X-ray absorption model \texttt{TBnew}\footnote{http://pulsar.sternwarte.uni-erlangen.de/wilms/research/tbabs/}. This model computes the total X-ray absorption cross section as the sum of the contributions from gas-phase material, dust grains, and molecules encountered by X-ray photons along the line of sight to the observer \citep{2000ApJ...542..914W}. The interstellar column density (ISM), $N_{\rm H}^{\rm ISM}$, was kept fixed at $N_{\rm H}^{\rm ISM} = 2.20 \times 10^{21}\,\mathrm{cm^{-2}}$, following \citet{2004ApJ...611..353C}, who derived this value from ultraviolet absorption measurements of \ion{H}{i} (IUE) and \ion{H}{2} (FUSE) reported by \citet{1994ApJS...93..211D} and \citet{2002ApJ...577..221R}, respectively. 
In addition to ISM absorption, we included an additional local absorber, characterized by a partial covering fraction (parameter \textit{CF}), which serves as an indicator of the degree of clumping in the donor star’s stellar wind.

\begin{figure*}
\centering
\subfigure{\includegraphics[trim={8cm 0cm 8cm 0cm},width=1\textwidth]{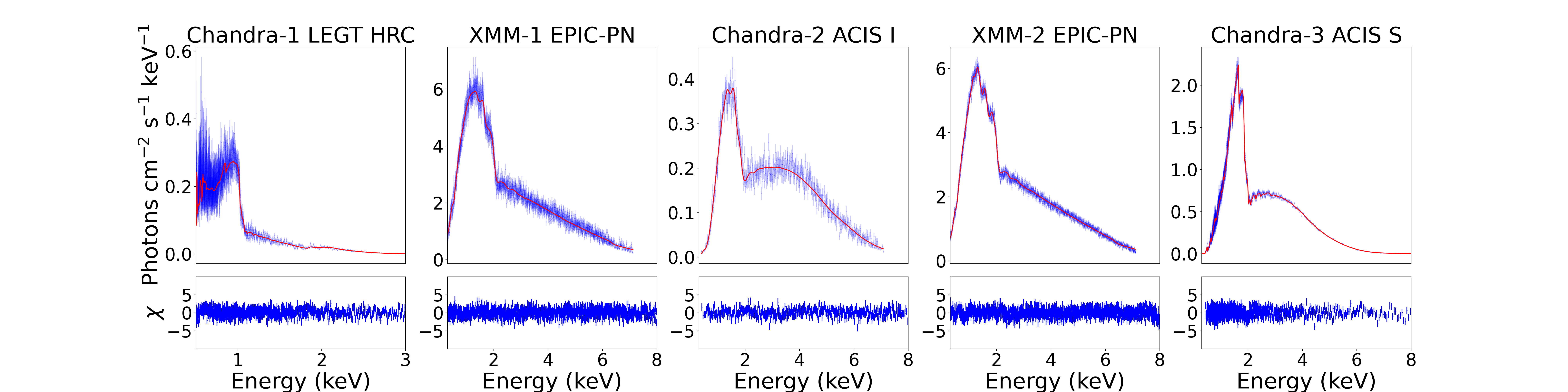}}
\subfigure{\includegraphics[trim={0cm 0cm 0cm 0cm},width=0.8\textwidth]{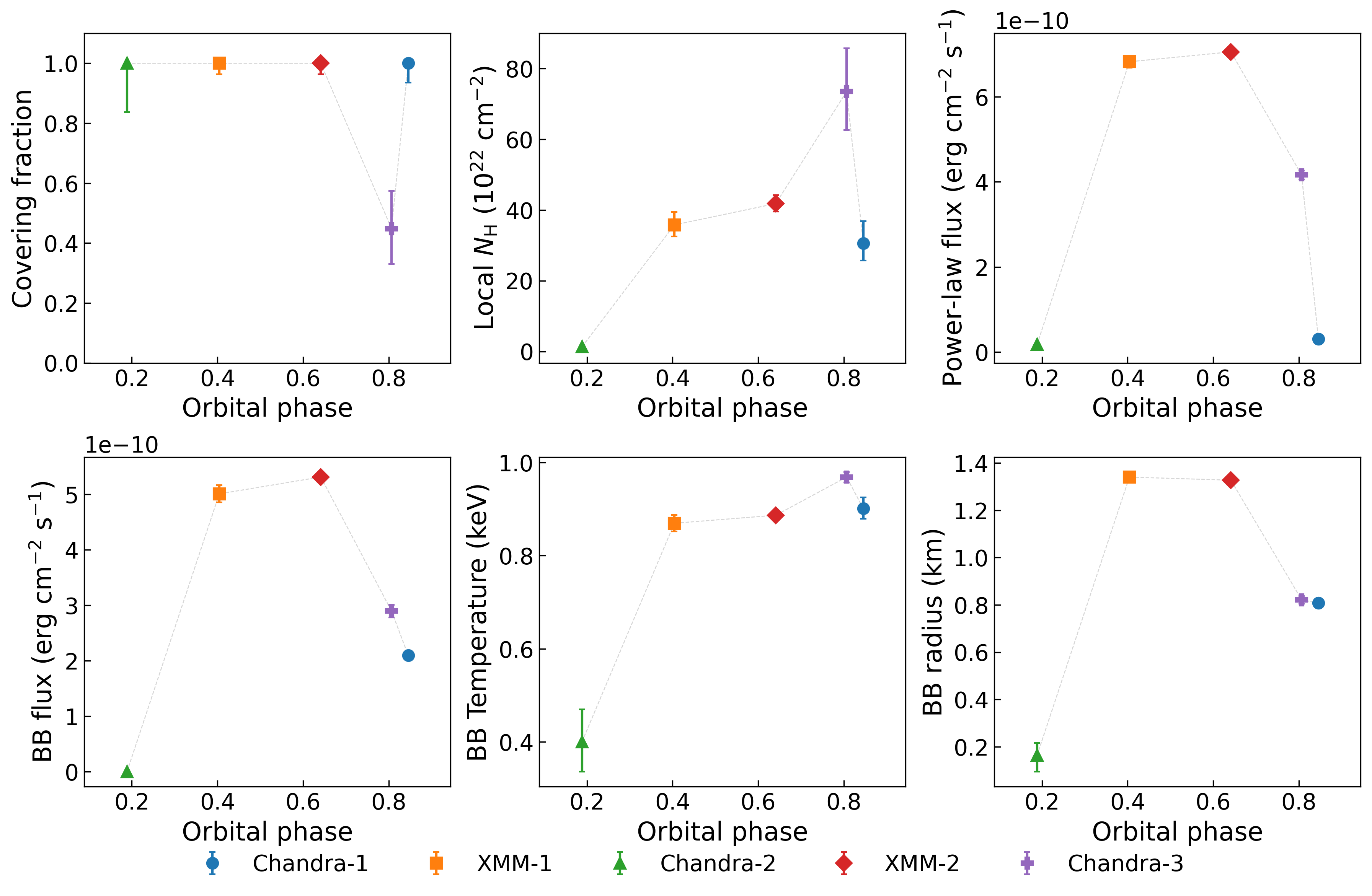}}
\caption{Upper panels: spectra, model and residuals of each observation. Lower panels: model parameters vs the orbital phase. Observation Chandra-1 is marked in red to emphasize that was taken with the LETG HRC instrument, and thus, covering different energy ranges.} 
\label{nh}%
\end{figure*}

To obtain statistically robust line significances, in the average spectra, following the recommendations of \citet{2002ApJ...571..545P}, we generated 1000 line-free \texttt{fakeit} spectra for each observation using the best-fitting continuum model and the corresponding instrument responses. Each simulated spectrum was processed with the same automated line-search procedure applied to the real data, and for every fake spectrum we recorded the maximum improvement in $\chi^{2}$ produced by noise fluctuations. This empirical null distribution provides the false-alarm probability for each candidate line. Features in the real spectra whose improvement in $\chi^{2}$ corresponded to a Monte-Carlo false-alarm probability below 1\% were considered statistically significant and were retained in the final model.
The model used is described by Eq. \ref{model}.

\begin{equation}
\label{model}
\begin{split}
 F(E) &= \left[ CF \exp\left(-N_{\rm H}^{\rm LOC} \sigma(E)\right) + \exp\left(-N_{\rm H}^{\rm ISM} \sigma(E)\right) \right]\\
& \quad \times \left[ \text{powerlaw} +\text{bbody} + \sum_{i=1}^{n} \text{Gaussian}_{i} \right],\\
\end{split}
\end{equation}
where $CF$ is the partial covering fraction ($0<CF<1$). The summation represents the emission lines present within the spectra. Each emission line was fitted with a single Gaussian function added to the continuum, all of which were modified by the absorption component. The spectra, best-fit model, and corresponding residuals are presented in the upper panel of Fig. \ref{nh}, while the evolution of the model parameters as a function of orbital phase is illustrated in the middle and lower panels of Fig. \ref{nh}. The continuum parameters are shown in Table \ref{tab:model_params}. Line parameters obtained for each average spectrum are collected in Table \ref{tab:appendixA}. The radii of the two \texttt{bbody} emitting regions can be estimated assuming an area $\pi R_{W}^{2}$, as shown in Eq. \ref{eq:radius} \citep{2004A&A...423..301T}

\begin{equation}
\label{eq:radius}
R_{W}=0.6\, \sqrt{L_{34}}\, (k\, T)^{-2} [\rm km],
\end{equation}
where $L_{34}$ is the luminosity in units of $10^{34}\,\mathrm{erg\,s^{-1}}$ and $k\, T$ is the temperature in keV. Their values along with the luminosity are collected in Table \ref{tab:derived_params}.

\begin{table*}
  \centering
  \caption{Best-fit spectral model parameters for each observation.}
  \begin{adjustbox}{max width=\textwidth}
    \begin{tabular}{lccccccc}
      \hline\hline
      \\
      Spectra & $\chi_{\rm r}^2$ & d.o.f. &
      $N_{\rm H}^{\rm LOC}$ & $K_{\rm pow}$ & $\Gamma$ &
      $K_{\rm bb}$ & $kT_{\rm bb}$ \\
      &&&
      ($\times10^{22}$ cm$^{-2}$) &
      photons keV$^{-1}$ cm$^{-2}$ s$^{-1}$ & &
      {$L_{39}/D^2_{10}$ } & keV \\
        &&&
       &
     ($\times10^{-3}$) & &
      ($\times10^{-3}$) &  \\
      \midrule

Chandra-1 & 1.1 & 1260 &
31$^{+6}_{-5}$ &
8.4$^{+0.7}_{-0.6}$ &
3.82$\pm$0.16 &
2.35$\pm$0.09 &
0.90$\pm$0.02 \\\\

XMM-1 & 1.1 & 1440 &
36$^{+4}_{-3}$ &
52.4$\pm$1.0 &
1.13$\pm$0.02 &
6.01$\pm$0.19 &
0.87$\pm$0.02 \\\\

Chandra-2 & 1.1 & 490 &
1.3$\pm$0.3 &
0.6$^{+0.1}_{-0.3}$ &
0.59$\pm$0.04 &
0.004$\pm$0.003 &
0.40$\pm$0.07 \\\\

XMM-2 & 1.2 & 1530 &
41.8$^{+2.4}_{-2.3}$ &
58.4$\pm$0.7 &
1.19$\pm$0.01 &
6.38$\pm$0.10 &
0.89$\pm$0.01 \\\\

Chandra-3 & 1.2 & 1340 &
74$^{+12}_{-11}$ &
27.5$\pm$0.7 &
1.01$\pm$0.03 &
3.5$\pm$0.1 &
0.97$\pm$0.01 \\

      \hline
    \end{tabular}
  \end{adjustbox}
  \label{tab:model_params}
\end{table*}

\begin{table}
  \centering
  \caption{Derived emission radii and luminosities for each observation.}
  \begin{adjustbox}{max width=\columnwidth}
    \begin{tabular}{ccc}
      \hline\hline
      \\
      Spectra & $R_{\rm bb}$ (km) & $L_{\rm X}$ ($10^{34}$ erg s$^{-1}$) \\
      \midrule

Chandra-1 & 0.81$\pm$0.02 & $\sim$1.2 \\\\

XMM-1     & 1.34$\pm$0.02 & $\sim$13 \\\\

Chandra-2 & 0.16$^{+0.05}_{-0.07}$ & $\sim$0.2 \\\\

XMM-2     & 1.33$\pm$0.01 & $\sim$14 \\\\

Chandra-3 & 0.82$\pm$0.02 & $\sim$5 \\

      \hline
    \end{tabular}
  \end{adjustbox}

  \label{tab:derived_params}
\end{table}

The normalization of the blackbody component is defined as $L_{39}/D_{10}^{2}$. Adopting a source distance of 0.69 kpc and using the luminosity derived for this component, we obtain values consistent with the normalizations inferred from the spectral fits, namely $2.5$, $6.0$, $0.004$, $6.4$, and $3.5$, all by a factor of $\times 10^{-3}$. This agreement supports the consistency of the adopted blackbody component and the assumed source distance (see Table \ref{tab:model_params} $K_{\rm bb2}$ column).

\subsection{NS  spin period}

\begin{figure*}
\centering
\subfigure{\includegraphics[trim={0cm 0cm 0cm 0cm},width=1\textwidth]{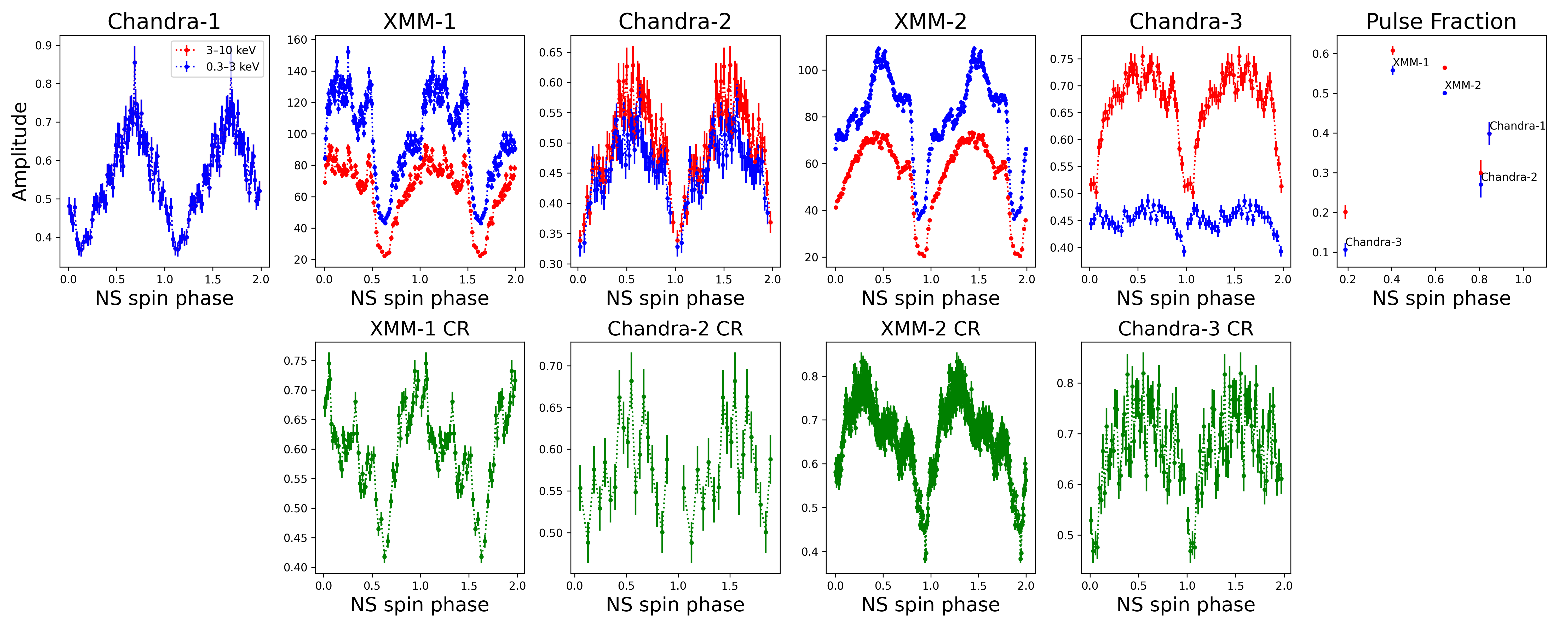}}
\caption{Upper row: Folded light curve to the NS spin period for each observation in the hard (red) and soft (blue) energy ranges. Observation Chandra-1 is only shown in the (0.3-3) keV energy range, as was obtained using the LETG HRC instrument, which has limited sensitivity to higher-energy photons and lacks spectral resolution. The right-most panel represents the pulse fraction on both the hard (red) and soft (blue) energy ranges vs the orbital phase for each observation. Bottom row: folded color ratio (CR) for each observation.} 
\label{fig:pulse_shape}%
\end{figure*}

\begin{figure*}
\centering
\subfigure{\includegraphics[trim={0cm 0cm 0cm 0cm},width=1\textwidth]{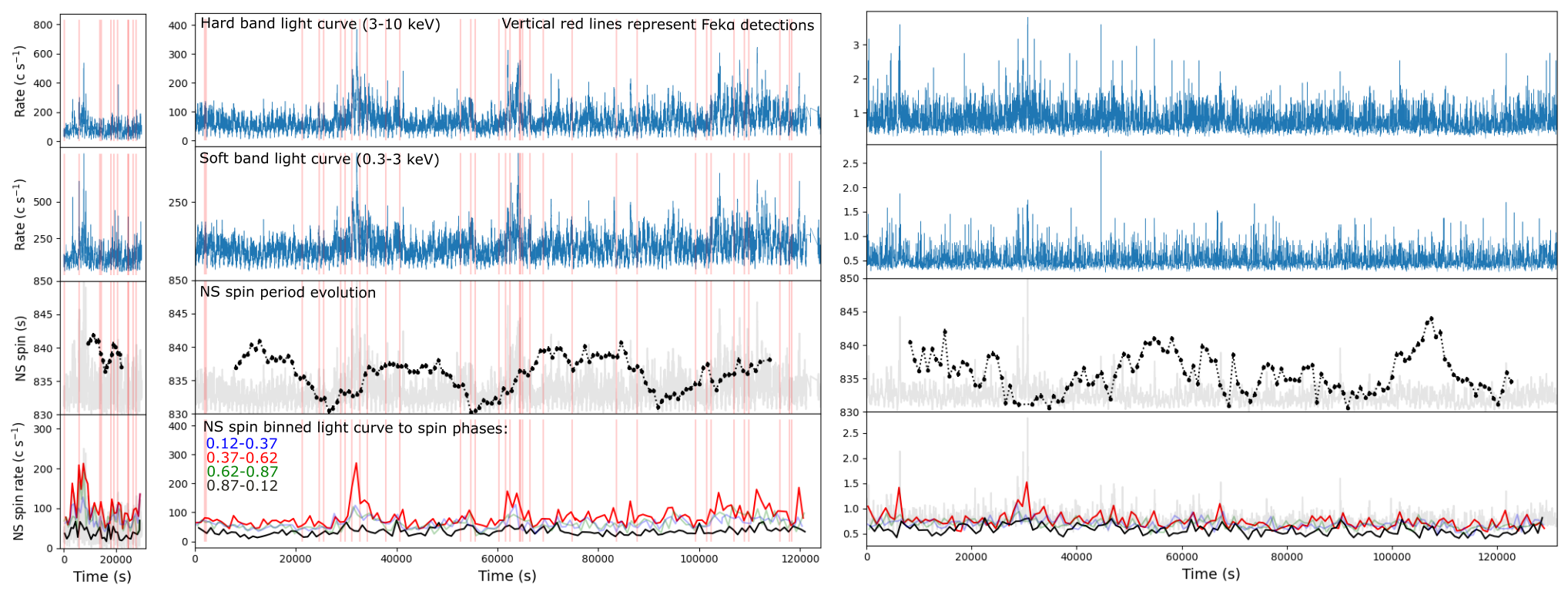}}
\caption{Light curves of XMM-1, XMM-2, and Chandra-3 (from left to right). Top panel: hard-band count rate (3--10)~keV. Second panel: soft-band count rate (0.3--3)~keV. Third panel: NS spin period as a function of time. Bottom panel: hard-band light curve binned to the NS spin phase intervals: 0.37--0.62 (pulse maximum; hot spot facing the observer; red) and 0.87--0.12 (pulse minimum; hot spot on the far side; black). Intermediate phase intervals are shown in light blue and green. Times of detected Fe~K$\alpha$ signatures are marked by vertical red lines. No Fe~K$\alpha$ signatures are detected in Chandra-3.}
\label{fig:phase_resolved}%
\end{figure*}

The NS spin period of this source is $\sim$ 837 s and can be distinguished by eye (Fig. \ref{fig:phase_resolved}, upper panel). To obtain the NS spin period of the light curves for each observation, we used a \textit{Lomb-Scargle} periodogram \citep{1976Ap&SS..39..447L,1982ApJ...263..835S}. We folded each  light curve for each observation and the color ratio (high energy light curve, $3-10$ keV, divided by low energy light curve $0.3-3$ keV) to the NS spin period and calculated its pulse fraction. The pulse fraction is a measure of the amplitude of variability in a periodic or pulsating signal. It can be calculated as shown in Eq. \ref{eq:pf}.

\begin{equation}
{PF} = \frac{I_{\text{max}} - I_{\text{min}}}{I_{\text{max}} + I_{\text{min}}},
\label{eq:pf}
\end{equation}
where $I_{\rm max}$ and $I_{\rm min}$ are the maximum and minimum intensity observed during the pulse, respectively. The results are collected in Table \ref{tab:pulse_log} and the folded pulses are shown in Fig. \ref{fig:pulse_shape}. 

\begin{table*}[ht]
\caption{ NS spin period results.}
\label{tab:pulse_log}
\centering
\begin{tabular}{lcccc}
\hline\hline
Observation & Period (s) & Pulse fraction (3–10 keV) & Pulse fraction (0.3–3 keV) & Color ratio \\
\hline
Chandra-1  & $835.9 \pm 0.5$ &$-$  &          $0.40 \pm 0.03$ & $-$      \\
XMM-1     & $837.1 \pm 0.4$ & $0.61 \pm 0.01$ & $0.56 \pm 0.01$ & $0.59 \pm 0.01$  \\
Chandra-2  & $835.7 \pm 0.6$ & $0.27 \pm 0.03$ & $0.26 \pm 0.03$ & $0.57 \pm 0.01$  \\
XMM-2  & $835.7 \pm 0.5$    & $0.56 \pm 0.01$ & $0.50 \pm 0.01$ & $0.65 \pm 0.01$  \\
Chandra-3  & $835.9 \pm 0.5$ & $0.20 \pm 0.02$ & $0.11 \pm 0.02$ & $0.65 \pm 0.01$  \\
\hline
\end{tabular}
\end{table*}

\begin{figure}
\centering
\subfigure{\includegraphics[trim={0cm 0cm 0cm 0cm},width=1\columnwidth]{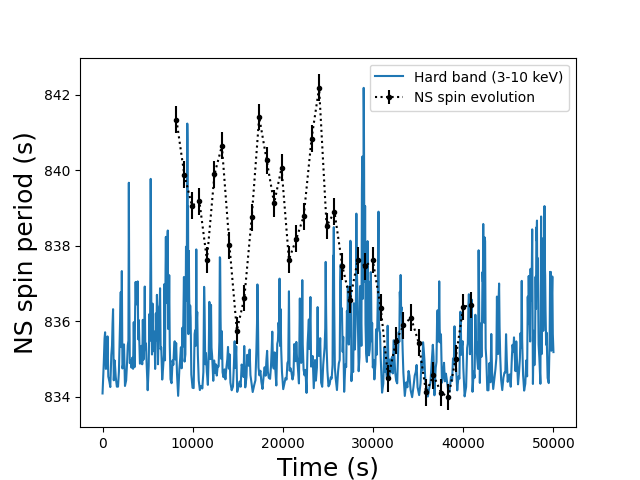}}
\caption{NS spin period as a function of time for observation Chandra-1. Black represents the NS spin evolution and blue the hard band (3-10 keV light curve).}
\label{fig:ns_spin_ev_chandra1.png}%
\end{figure}

 For observations Chandra-1, XMM-1, XMM-2, and Chandra-3, the evolution of the NS spin period was measured using a sliding-window analysis, where each window contained roughly 20 NS spin cycles (837$\times$ 20 s) and was shifted forward by one NS spin period (837 s) at each step, keeping those that showed peaks within the \textit{Lomb-Scargle} periodograms with at least a signal-to-noise ratio of 10 and a false alarm probability smaller than $10 ^{-4}$. The results obtained are collected on the third panel in Fig. \ref{fig:phase_resolved} for XMM-1, XMM-2 and Chandra-3 and on Fig. \ref{fig:ns_spin_ev_chandra1.png} for Chandra-1. The spin period shows variation around the equilibrium value P$_0\sim$ 836 s with a characteristic time of  $\tau\approx 3 \times 10^4$ s. This method was also applied in the analysis of Chandra-2 observations. However, the X-ray flux was 60–20 times lower than in the other datasets and no pulses meeting our criteria were found.

\subsection{NS spin resolved spectral analysis.}
\label{description_phase_resolved}

X Persei is a very bright source, which allows us to perform a detailed phase resolved analysis. Instead of simply dividing the lightcurve in sections of enough S/N ratio, we performed NS spin resolved spectral analysis. This is possible for not only the brightness of the source but also its slow pulsation. This resolved spectral analysis refers to the NS spin around its axis and is not related to the orbital period. 

Therefore, we divided each observation into four NS spin phases: $0.12-0.37$, $0.37-0.62$, $0.62-0.87$, and $0.87-0.12$. These segments correspond approximately to the rise, maximum (hot spot towards the observer), decrease, and minimum (hot spot opposed to observer) phases of the NS spin cycle, each spanning about 210 seconds.

\subsubsection{Continuum components as a function of the NS spin phase}
First, for each observation we combined the spectra corresponding to each NS spin segment. This allows us to observe the evolution of the continuum components with the NS spin phase. The results obtained for each fit are collected on Table \ref{tab:spin_resolved_params}. 

For each continuum component we calculated the pulse fraction of their normalizations following Eq. \ref{eq:pf}, taking NS spin phase $0.37-0.62$ as $I_{max}$ and $0.87-0.12$ as $I_{min}$.

In general, the powerlaw and blackbody components are pulsating according to the NS spin in every observation, showing its maximum when the hot spot is oriented towards the observer. 

\begin{figure}
\centering
\subfigure{\includegraphics[trim={0cm 0cm 0cm 0cm},width=1\columnwidth]{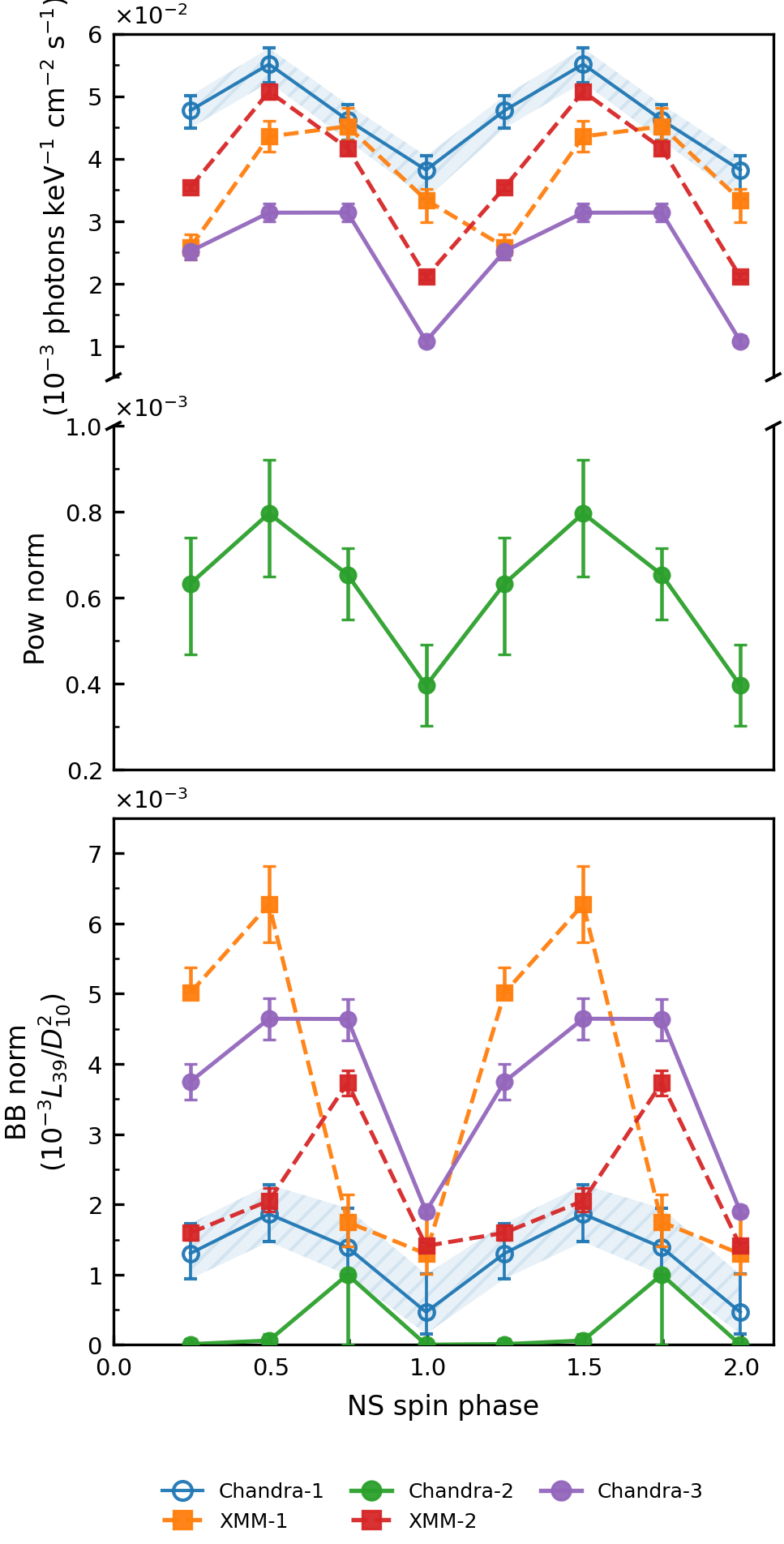}}

\caption{Spin-phase–resolved evolution of the continuum normalizations of X Persei. The powerlaw (upper) and blackbody (bottom) normalizations are shown as a function of NS spin phase and folded over two cycles for clarity. For each component, the same data are displayed in two vertically attached panels with different y-axis ranges to highlight both large- and small-amplitude variations. The powerlaw normalization has units of $\mathrm{photons}\,\mathrm{keV}^{-1}\,\mathrm{cm}^{-2}\,\mathrm{s}^{-1}$, while the blackbody normalization is given in units of $L_{39}\, D_{10}^{-2}$, where $D_{10}$ is the distance to the source in units of 10 kpc and $L_{39}$ is the luminosity in units of $10^{39}$ erg s$^{-1}$. \textit{Chandra} observations are shown in blue and green, and \textit{XMM-Newton} observations in red and orange.} 
\label{fig:cont_resolved}%
\end{figure}

\begin{table}
\centering
\caption{Pulse fraction of the powerlaw and blackbody continuum components. }

\begin{tabular}{lcc}
\hline
Observation & Powerlaw & Blackbody \\
\hline
Chandra-1  & $0.2 \pm 0.1$ & $0.6 \pm 0.6$ \\
XMM-1      & $0.13 \pm 0.08$ & $0.7 \pm 1.4$ \\
Chandra-2  & $0.3 \pm 0.3$ & $0.9 \pm 0.9$ \\
XMM-2      & $0.4 \pm 0.1$ & $0.19 \pm 0.01$ \\
Chandra-3  & $0.5 \pm 0.1$ & $0.4 \pm 0.1$ \\
\hline
\end{tabular}

\tablefoot{The reported uncertainties were estimated from the confidence intervals of the maximum and minimum phase bins used to compute the pulse fraction.}

\label{table:pf_comps}
\end{table}

\subsubsection{Fe K$\alpha$ emission features in NS spin resolved spectra}
Although the X Persei spectrum shows few emission lines, we proposed that certain features might arise only intermittently and therefore be washed out in the phase-averaged spectrum. In particular, we focus on the Fe K$\alpha$ emission, which is a well-known signature of cold and dense stellar-wind material, making it a useful tracer of the presence of clumps within the disk. To investigate this, we searched for transient emission line features by analyzing spectra resolved over the NS spin phase, using time bins of 210 s. This approach yielded reliable results for the \textit{XMM-Newton} observations (XMM-1 and XMM-2) and for Chandra-3, but could not be applied to observations Chandra-1 and Chandra-2 (see Table \ref{tab:obs_usage}).

\begin{figure}
\centering
\subfigure{\includegraphics[trim={0cm 0cm 0cm 0cm},width=1\columnwidth]{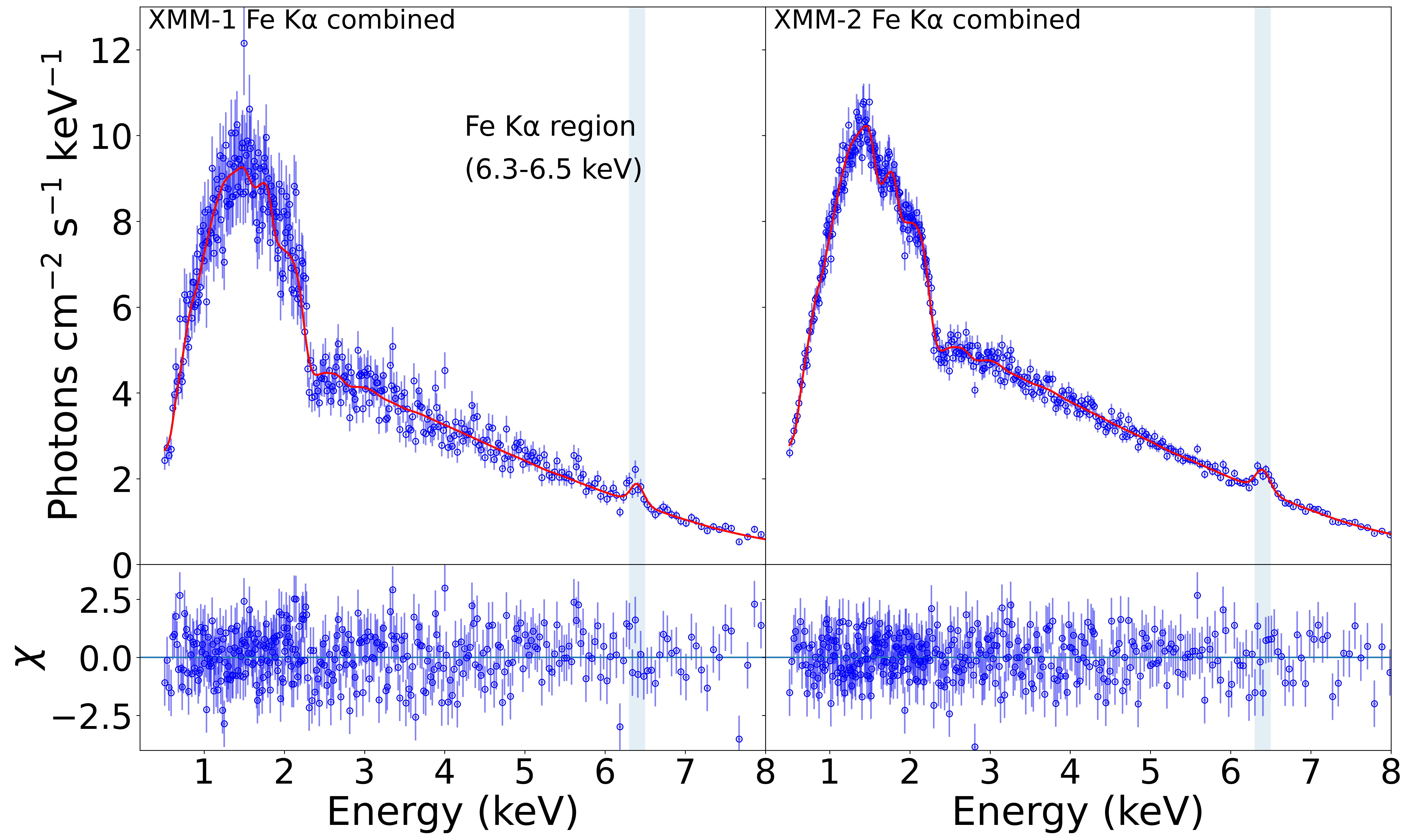}}
\caption{Combined Fe K$\alpha$ spectra obtained from the XMM–1 and XMM–2 observations. The upper panels display the observed spectra (blue) along with the best-fitting spectral model (red), whereas the lower panels present the corresponding fit residuals in terms of $\chi$. The best-fit parameters are summarized in Table \ref{table:feka_values}.}
\label{fig:fekafootprint}
\end{figure}

{To quantify the statistical significance of the Fe~K$\alpha$ detections we estimate the false-positive rate of our automated line-search pipeline using \texttt{fakeit} simulations that contain no Fe~K$\alpha$ line. In both the real and simulated spectra we run the same blind line search over the full energy range and classify a spectrum as an Fe~K$\alpha$ detection when at least one candidate line falls within a pre-defined window around the expected Fe~K$\alpha$ energy, $E=(6.4\pm0.1)$~keV. We define the per-spectrum false-alarm probability as $p_0 \equiv N_{\rm det}^{\rm (fake)}/N_{\rm tot}^{\rm (fake)}$, where $N_{\rm det}^{\rm (fake)}$ is the number of simulated spectra that yield at least one candidate line in the Fe~K$\alpha$ window and $N_{\rm tot}^{\rm (fake)}$ is the total number of simulated spectra. Assuming independent trials across the $N_{\rm tot}$ real spectra, the global (multi-trial) probability of obtaining at least $N_{\rm det}$ detections under the null hypothesis is given by the binomial upper tail,
\begin{equation}
p \equiv P(X \ge N_{\rm det}\,|\,N_{\rm tot},p_0) = \sum_{k=N_{\rm det}}^{N_{\rm tot}} \binom{N_{\rm tot}}{k} p_0^k (1-p_0)^{N_{\rm tot}-k}.
\end{equation}
To assess the significance of the Fe K$\alpha$ features, we performed \texttt{fakeit} simulations for each dataset, obtaining 140 simulated spectra for XMM-1 and 400 for XMM-2. Note that these simulations are distinct from the 1000 line-free \texttt{fakeit} spectra used in the average-spectrum line search described above. Here we use dedicated \texttt{fakeit} simulations tailored to the NS spin-resolved Fe K$\alpha$ analysis. Within the adopted Fe K$\alpha$ energy interval, we found 11 and 35 features in the real XMM-1 and XMM-2 spectra, respectively, compared with 5 and 3 features in the simulated spectra. The corresponding false-positive rates estimated from the simulations are $p_0 = 0.036$ for XMM-1 and $p_0 = 0.012$ for XMM-2, yielding binomial tail probabilities of $p = 7.5 \times 10^{-3}$ and $p = 7 \times 10^{-26}$, respectively.

The transient nature of the line is consistent with an origin in sparse clumps illuminated by the NS as they are accreted. Recombination times for typical clump densities, $n_{e} \sim 10^{(10-14)}~\mathrm{cm^{-3}}$, range from $t_{\rm rec} \sim (0.01-30)$  s and are therefore diluted in the averaged spectra, while still occasionally detectable in the 210~s spectra. 

We combined all time resolved spectra containing Fe K$\alpha$ signatures from the XMM–1 and XMM–2 observations, respectively. Table \ref{table:feka_values} summarizes the combined Fe K$\alpha$ spectral fits. Fig. \ref{fig:fekafootprint} shows the Fe K$\alpha$ combined spectra for XMM-1 and XMM-2.

\begin{table}
\centering
\caption {Best-fit Fe K$\alpha$ emission-line parameters obtained from the combined Fe K$\alpha$ time-selected spectra (see Fig.~\ref{fig:fekafootprint}). }
\begin{adjustbox}{max width=\columnwidth}
\begin{tabular}{lrr}

Combined spectra Fe K$\alpha$ fit values &XMM-1&XMM-2\\
\hline
Centroid (keV)&$6.40^{+0.02}_{-0.07}$ &$6.42\pm0.02$\\
EW (eV)& 70 $\pm$ 10 &	70 $\pm$ 30 \\
Flux $\times$ 10$^{-4}$ (photons s$^{-1}$ cm$^{-2}$) & $3.9^{+1.5}_{-1.3}$  & $3.7^{+0.7}_{-0.6}$ \\
Prevalence & 8\% &	9\% \\
\hline     
\\
\end{tabular}
\end{adjustbox}
\label{table:feka_values}
\end{table}

\subsection{Features in light curves}

X-ray light curves present sudden increases and decreases in their flux in different energy ranges, on time scales much shorter than the orbital and spin periods, called spikes and dips, respectively. We analyze these features and associate them with different astrophysical phenomena.

\subsubsection{Dips as clump candidates}

X-ray photons emitted by the source traverse a complex environment before reaching the telescope, including the circumstellar medium and dense, inhomogeneous structures such as cold and dense regions known as clumps. When X-ray emission passes through such an over-dense material, it can produce partial occultation and enhanced absorption at low energies, leading to dips in the light curves. We will consider as clump candidates those dips that are simultaneously present in both soft and hard energy light curves, overlapping for at least 50\% of their duration, and are more pronounced in the low-energy band, where absorption effects are stronger. This method was used successfully in previous analysis \citep{2024A&A...690A.360S}. We identified 20, 3, 203, and 0 high-probability clump candidates in XMM-1, Chandra-2, XMM-2, and Chandra-3, respectively. Here, high-probability candidates are those assigned a probability higher than 90\% by the \texttt{dipspeaks} algorithm. These correspond to occurrence rates of 3, 0.2, 6, and 0 clumps~h$^{-1}$, respectively.

\subsubsection{Spikes as inhomogeneous accretion processes}

On the other hand, the accretion process itself is highly variable and non-uniform. The luminosity of the source is linked to the mass accretion rate as shown in Eq. \ref{eq:0}.

\begin{equation} 
L = \frac{G M \dot{M}}{R},
\label{eq:0}
\end{equation}
where $L$ is the X-ray luminosity, $G$ is the gravitational constant, $M$ is the mass of the NS, $\dot{M}$ is the mass accretion rate and $R$ is the radius of the NS. Sudden increases in luminosity (observed as spikes) are often associated with Rayleigh–Taylor instabilities (RTIs). In this scenario, the entry rate of matter leads to a short and sudden increase of $\dot{M}$ and thus to the luminosity. 

 To detect RTI signatures we searched for spikes in the hard energy range lightcurve ($3-10$ keV), as it is less affected by absorption processes. RTIs detected in this manner were reported in \citet{2025A&A...694A.192S}. In the present work, we found 42, 10, 310 and 32 high probability spikes (those assigned a probability higher than 90\% by the \texttt{dipspeaks} algorithm) in XMM-1, Chandra-2, XMM-2 and Chandra-3 observations, respectively, pointing to a rate of 5, 0.7, 9, 1 spikes ~h$^{-1}$.

\section{Discussion}

\subsection{The disk }
\label{sec:disc}

The density structure of the BeXB decretion disk can be described with the viscous decretion disk model \citep{2017A&A...601A..74K}. In cylindrical coordinates $(r,z)$ centered on the donor star, the density is given by
\begin{equation}
\rho(r,z) = \rho_0 \left(\frac{r}{R_{\rm e}}\right)^{-\alpha}
\exp\!\left(-\frac{z^2}{2H^2}\right),
\label{eq:rho}
\end{equation}
where $r$ is the distance from the donor star in the disk mid-plane and $z$ is the height above the disk plane, $\rho_0$ is the base density at the stellar surface, $R_{\rm e}$ is the equatorial radius of the rotationally deformed star, $\alpha$ is the radial density exponent, and $H$ is the disk scale height. The radial dependence of the scale height is
\begin{equation}
H(r) = H_0 \left(\frac{r}{R_{\rm e}}\right)^{3/2},
\label{eq:Hofr}
\end{equation}
with the scale height at the disk base given by
\begin{equation}
H_0 = \frac{c_s R_{\rm e}}{v_{\rm orb}},
\label{eq:H0}
\end{equation}
where $c_s$ is the sound speed and $v_{\rm orb}$ is the orbital velocity of the disk material.

Most Be stars undergo complex phases of disk build-up and dissipation. In their models \cite{2007ASPC..361.....O}, values of $\alpha$ < 3.5 are only reached when the disk is in dissipation, while $\alpha$ > 3.5 are attributed to disks in the build-up stage.

The material that X-ray photons must traverse on their way to the observer is governed by several factors. These are the orbital elements of the system: orbital separation, eccentricity, and argument for periapsis. $N_{\rm H}$ also varies with the orientation of the system with respect to the observer, the orbital inclination, and the orbital phase. Because inclination and eccentricity modulate the line-of-sight path length, they produce phase-dependent changes in the observed column density. In the special limit of zero inclination and zero eccentricity, no orbital modulation would be expected. Using Eqs. \ref{eq:rho}, \ref{eq:Hofr} and \ref{eq:H0}, together with the system parameters listed in Table \ref{parameters}, we created a 3D density map and integrated the density that the X-ray emitted radiation from the NS must cross along its path to the observer, as proxy for the local absorption column $N_{\rm H}^{\rm LOC}$. In Fig. \ref{fig:bex_deg}, we present this three-dimensional model. The color gradient illustrates the spatial variations in disk density, while the dashed arrows indicate two representative examples of the density integration paths used to compute \(N_{\mathrm{H}}\). We then compare this theoretically derived $N_{\rm H}$ with the value obtained from our spectral fitting, with the goal of constraining the disk structure, specifically the density powerlaw index $\alpha$ and the disk density near the donor star, $\rho_{0}$.

The complexity of the model prevents us from using a classical least-squares approach; hence, the data were fitted using a particle swarm optimization (PSO) algorithm. In PSO, each particle represents a potential solution to an optimization problem. The particles move through the solution space allowing to find the optimal parameter configuration \citep{10.1162/EVCO_r_00180}.

From optical and IR photometry, \citet{1998MNRAS.296..785T} inferred a high inner-disk (photospheric) base density of $\rho_{0}\simeq 1.5\times10^{-10}\,\mathrm{g\,cm^{-3}}$, varying by at least a factor of $\sim 20$ between optical high and low states. We adopt this range as a prior and therefore allow $\rho_{0}$ to vary between $10^{-11}$ and $10^{-8}$~$\mathrm{g\,cm^{-3}}$, while the radial powerlaw index is allowed to vary within $2\le \alpha \le 5$. After 10 iterations for each observation, we find that, as expected, $\alpha$ and $\log_{10}\rho_{0}$ are strongly degenerated. We found that $\alpha$ varies from $2.3-3.2$ and $\rho_{0}$ ranges from $(6.5-2)\times10^{-10}\,\mathrm{g\,cm^{-3}}$. \citet{1998MNRAS.296..785T} instead derived a steep radial exponent, $\alpha \approx 5$. A direct comparison of absolute values should be made with caution, since \citet{1998MNRAS.296..785T} assumed a nearly pole-on disk, while we adopted an inclination of $30^\circ$. Overall, both studies point to a dense inner disk and a relatively compact density distribution. Interestingly, in the case of Chandra-2, which corresponds to the lowest luminosity state in our sample, the disk shows the highest value of the $\alpha$ parameter, which may indicate a more compact configuration compared to the other observations.

\begin{table}
\centering
\caption{Best-fit parameters of the Be-disk absorption model for each observation.}
\begin{tabular}{lccc}
\hline
Observation & $N_{\rm H}^{\rm LOC}$ & $\alpha$ & $\log_{10}\rho_{0}$  \\
 & ($\times10^{22}$ cm$^{-2}$) &  & (g cm$^{-3}$) \\
\hline
Chandra-1 & 31$^{+6}_{-5}$             & $2.4 \pm 0.5$ & $-9.2 \pm 0.8$ \\
XMM-1     & 36$^{+4}_{-3}$              & $2.5 \pm 0.6$ & $-9.1 \pm 1.0$ \\
Chandra-2 & $1.3 \pm 0.3$    & $3.3 \pm 0.5$ & $-9.2 \pm 0.9$ \\
XMM-2     & $42\pm 2$   & $2.6 \pm 0.6$ & $-9.0 \pm 1.0$ \\
Chandra-3 & $74^{+12}_{-11}$       & $2.5 \pm 0.4$ & $-8.7 \pm 0.7$ \\
\hline
\end{tabular}
\tablefoot{$N_{\rm H}^{\rm LOC}$ is the local equivalent hydrogen column density, $\alpha$ is the
radial density exponent, and $\rho_{0}$ is the density normalization.
Uncertainties on $\alpha$ and $\rho_{0}$ correspond to their standard deviations.}
\label{table:disk_parameters_ps}
\end{table}

\begin{figure}
\centering
\subfigure{\includegraphics[trim={0cm 0cm 0cm 0cm},width=1\columnwidth]{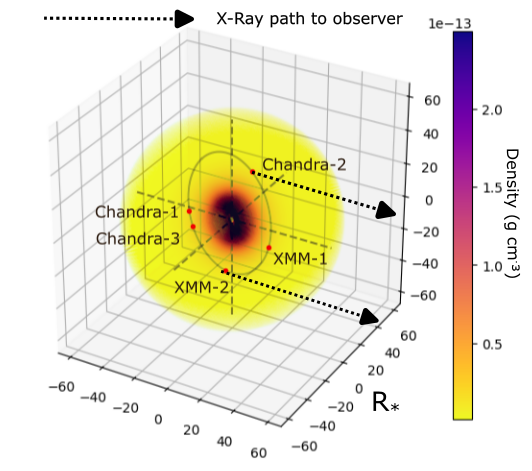}}
\caption{
Three-dimensional view of the BeXB circumstellar disk and the locations of the observations along the NS orbit. The color scale gradient represents the density, used to compute $N_{\mathrm{H}}$, for the adopted geometry and viewing angle. Dashed arrows represent two examples of the integration density path to calculate NH. Dimensions are expressed in units of the stellar radius, R$_{\star}$.
}
\label{fig:bex_deg}
\end{figure}

\begin{figure*}
\centering
\subfigure{\includegraphics[trim={0cm 0cm 0cm 0cm},width=1\textwidth]{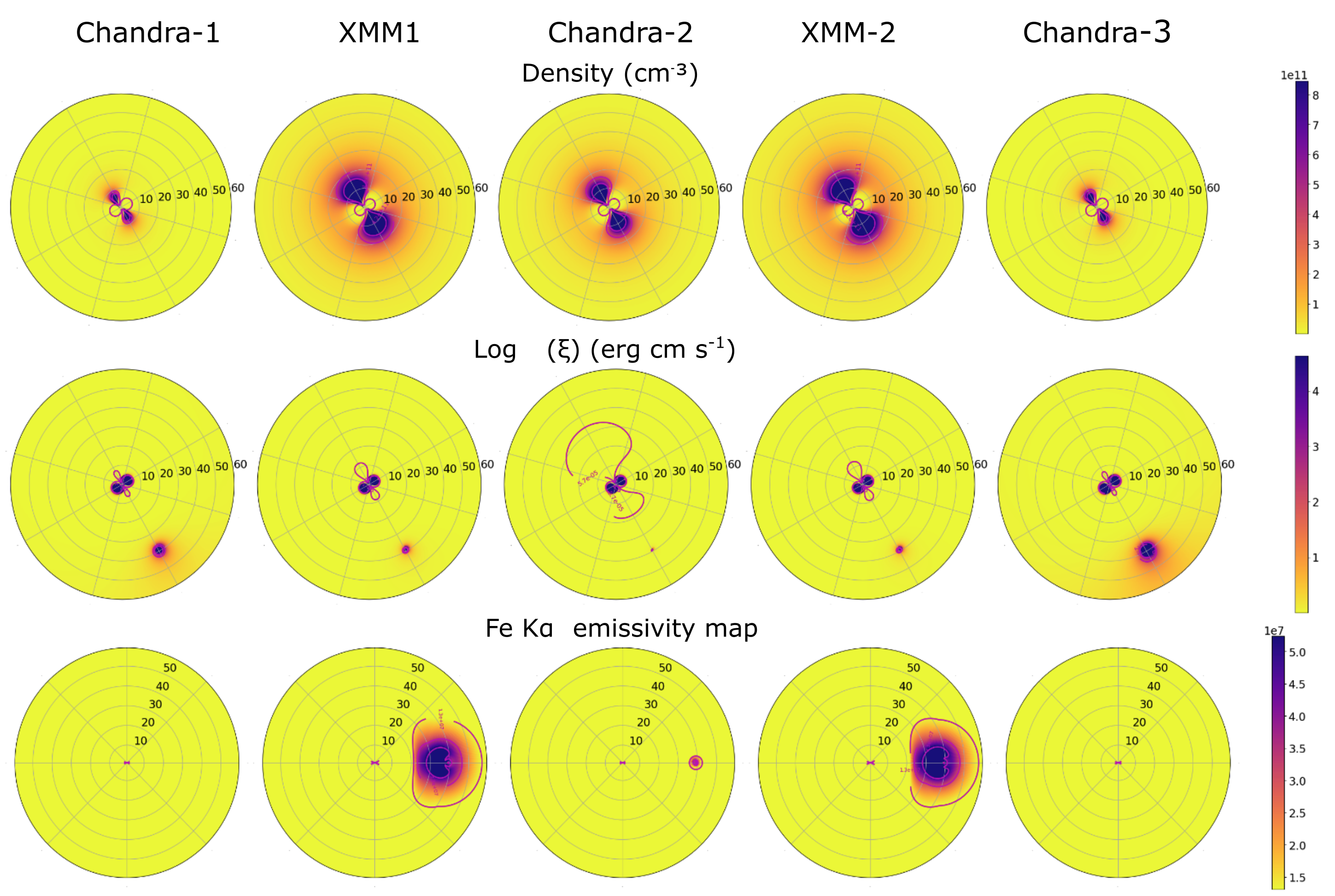}}
\caption{From the top down, first row: side-view density distribution for each observation, using the particle swarm derived parameters (see Table \ref{table:disk_parameters_ps}). Second row: Ionization parameter, calculated through the density profile and luminosity found for each observation. Third row: Fe K$\alpha$ reconstructed emission map from the density and ionization map. The radial axis is scaled in R$_{\star}$ units.}  
\label{fig:fe_emission}%
\end{figure*}

The near-neutral Fe K$\alpha$ emission line, typically formed in dense and cold environments, is a prominent feature in many high-mass X-ray binaries (HMXBs) with supergiant donors. These systems host compact objects orbiting at small separations $\sim (1.5-4)\,R_\star$ (stellar radius), where the stellar wind density is high and the compact object is well embedded in the wind. In contrast, X~Persei has a much wider orbit, of $\sim$ 40 $R_\star$, resulting in a significantly lower wind density at the NS location ($\sim$ 10$\times 10^{-18}$ g cm$^{-3}$ assuming a generic 1.5$\times 10^{-7}$ $M_\odot$ yr$^{-1}$ mass loss rate), and, as expected, no Fe K$\alpha$ emission is detected in phase-averaged spectra. As we have shown, however, it is detectable during short periods of time in some NS spin resolved spectra. In contrast, the density of the disk at the location of the NS is expected to be of the order of 10$^{-15}$ g cm$^{-3}$, approximately three orders of magnitude higher, with the disk being the primary reservoir of matter to be accreted by the NS.

 Using the derived disk parameters $\alpha$ and $\rho_0$ (Table~\ref{table:disk_parameters_ps}), the luminosities obtained from the spectral fits (Table~\ref{tab:derived_params}), and the system’s orbital parameters (Table~\ref{parameters}), we reconstructed the system’s density and ionization structure. This allows us to identify the regions of the circumstellar environment that meet the conditions for Fe K$\alpha$ emission. The emissivity of the Fe K$\alpha$ line depends on the local iron abundance, gas density, and ionization parameter $\xi$. To quantify this dependence, we used the photoionization code \texttt{XSTAR} \citep{2021Atoms...9...12M} to model the Fe K$\alpha$ emissivity as a function of $\log\xi$ over the range ($-2,5$ erg cm s$^{-1}$) for a thin plasma. This approach allowed us to identify the spatial regions where Fe K$\alpha$ emission is expected to form. Although this model represents a simplified description of the system geometry, neglects time-dependent ionization effects and assumes a smooth disk, it provides a first-order, qualitative assessment of whether and where in the system the required conditions for Fe K$\alpha$ emission are met. 

Searches for emission-line signatures in the NS spin-resolved spectra were only feasible for XMM-1, XMM-2, and Chandra-3, which makes a comparison among these observations particularly relevant. As shown in the third panel of Fig.~\ref{fig:fe_emission}, XMM-1 and XMM-2 exhibit extended regions consistent with Fe K$\alpha$ emission. Conversely, and in agreement with our observations, no Fe K$\alpha$ emission is expected in Chandra-3.

The presence of Fe~K$\alpha$ emission features in XMM-1 and XMM-2 is accompanied by the detection of clump candidates, whereas neither clump candidates nor Fe~K$\alpha$ emission features are detected in Chandra-3. Consistently, the CF, used as a proxy for the degree of clumping of circumstellar environment, reaches its lowest value for the Chandra-3 observation among the observations analyzed in this work. We therefore propose that the clumpier environments inferred for XMM-1 and XMM-2, together with their higher luminosities and access to denser inner regions of the circumstellar wind, naturally account for the transient nature of the Fe~K$\alpha$ emission. In such conditions, over-dense clumps are expected to be rapidly ionized when exposed to the intense NS radiation field, leading to short-lived features.

\subsection{The X-ray emission and NS spin}
\label{sec:x-rayemission}

\begin{figure}

\centering
\subfigure{\includegraphics[trim={0cm 0cm 0cm 0cm},width=1\columnwidth]{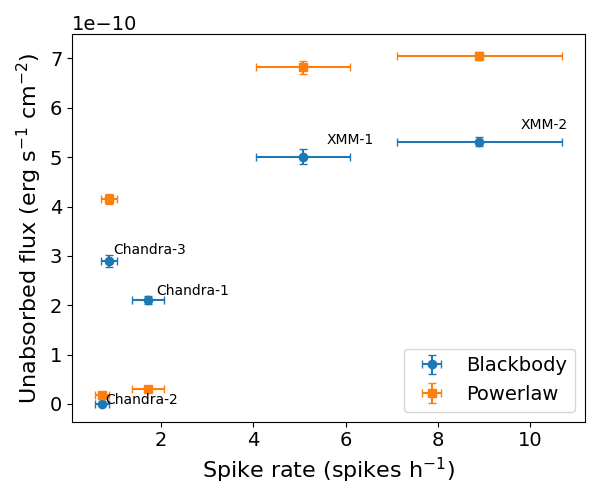}}
\caption{Powerlaw and blackbody unabsorbed fluxes as a function of the spike rate.}
 
\label{fig:density_2}
\end{figure}

X Persei exhibits some orbital modulations, consistent with the NS rotation around the donor star in an inclined and slightly eccentric orbit. The unabsorbed fluxes of all three spectral components are enhanced near orbital phases $0.4-0.6$, when the NS is closer to the observer. 
Observations XMM-1 and XMM-2 exhibit a high-soft state (high luminosity and softer emission), while \textit{Chandra} observations were taken with the source in a hard-low state (lower luminosity and harder emission). This can be appreciated both in the spectral parameters (Table \ref{tab:model_params} and Table \ref{tab:derived_params}), and in the folded pulses (Fig. \ref{fig:pulse_shape}).

The spike rate, i.e. short-lived enhancements in the X-ray light curves, increases monotonically with the fluxes of the pulsating continuum components ( blackbody and powerlaw; see Fig.~\ref{fig:density_2}), although the limited number of observations prevents us from establishing a statistically robust correlation.

Our hypothesis is that the observed spikes could be generated by irregular accretion of matter entering the magnetosphere as RTI. This phenomenology can be explained within the quasi‑spherical settling‐accretion paradigm \citep{2012MNRAS.420..216S}. In slowly rotating accreting X-ray pulsars with moderate X-ray luminosity (below $\sim 4 \times 10^{36}$ erg s$^{-1}$), a hot convective quasi-spherical shell can form above the NS magnetosphere. The matter enters the magnetosphere through RTI at the magnetospheric boundary via Compton cooling or radiative losses. This accretion process in HMXBs has been known and studied for a long time \citep[see][for instance]{1976ApJ...207..914A,1977ApJ...215..897E}. For a magnetic field of the order of $\sim10^{14}\,\mathrm{G}$, the Alfvén radius is estimated to be of the order of $ \sim10^{10}\,\mathrm{cm}$.
 
A striking feature supporting the quasi-spherical accretion frame is an almost sine-like spin period variation around the equilibrium value P$_0\sim$ 836 s (Fig. \ref{fig:phase_resolved} third panel). The characteristic time is $\tau\approx 3 \times 10^4$ s clearly visible in both XMM-2 and Chandra-3 data.  It means that the NS spin period regularly alternates on this timescale with the characteristic spin-up/down value $\dot P \sim 10/\left(1.5\times 10^4\right) \sim 10^{-3}$ where \(\dot P \equiv dP/dt\) is the time derivative of the spin period \(P\), expressed in s\,s\(^{-1}\), corresponding to the specific torque $I\dot\omega \sim \pm 2\, \pi\, 10^{-9}$, which has units of g\,cm\(^2\)\,s\(^{-2}\), where $\omega=2\, \pi/P$ is the NS spin frequency and $I$ is the NS moment of inertia. This is exactly what is expected in the quasi-spherical settling accretion with an Alfvén radius around $R_A=10^{10}$ cm and an accretion rate $\dot M\sim 10^{14}$ g\,s$^{-1}$ onto equilibrium low-luminosity X-ray pulsars.
The explanation of the regular torque reversal around the equilibrium NS spin period on the time scale $\tau=3\times 10^4$ s can be as follows.
The spin-up/down occurs due to the imbalance (positive/negative) between the angular velocity of the accreting matter $\omega_m$ and that of the NS magnetosphere at the Alfvén radius (see e.g. \cite{2012MNRAS.420..216S} for more detail) . The imbalance keeps for the time the magnetic field at $R_A$ needs to admix with the accreting matter, which is the magnetic diffusion time in the turbulized plasma:
$$
\tau_d= R_A^2/\eta_m,
$$
where $\eta_m$ is the magnetic diffusion coefficient. On this timescale, the toroidal magnetic field component, arising due to the $\omega_m - \omega$ angular velocity shear, saturates from the magnetospheric poloidal one to the maximum possible value $B_t\sim B_p$,  where $B_t$ is the toroidal magnetic-field component and $B_p$ is the poloidal magnetic field component \citep{1995MNRAS.275..244L,2012MNRAS.420..216S}. The magnetic diffusion coefficient can be written as $\eta_m\sim \nu_t$, where $\nu_t$ is the turbulent viscosity coefficient, $\nu_t\sim u_t l_{t}$. In our problem, the characteristic size of the largest turbulent eddy in the shell around the rotating magnetosphere is $l_t\sim \kappa R_A$, where $\kappa <1$ is the characteristic amplitude of inhomogeneity that initiates turbulence (for example, for a rotating sphere $\kappa=0$, for a dipole-like magnetosphere $\kappa\approx 0.3$). The characteristic turbulent velocity in this region should be of the order of the velocity of matter  penetrating through the magnetosphere due to the RTI $u_{ff}f(u)$. Therefore, we arrive at the estimate
$$
\tau_d =R_A^2/\eta_m  \sim R_A^2/(\nu_t l_t) = R_A/(u_{ff}f(u)\kappa) = t_{RT}/(f(u)\kappa)
$$
Here u$_{ff}$ is the free-fall velocity at the magnetospheric radius and $f(u)<1$ is the settling accretion factor such that $u_{ff}f(u)$ is the accreting matter penetration velocity through the magnetosphere \citep{2012MNRAS.420..216S}. With the moderate X-ray luminosity $L_x\lesssim 10^{36}$ erg/s and radiative plasma cooling $f(u)\sim 0.1$ we expect $\tau_d\sim 30\, t_{RT}$. The RTI time $t_{RT}$ can be read off the duration of spikes on the X-ray light curve. The observed spike durations are widely distributed, ranging between 120 s and 1600 s and an average of 600 s. Assuming $t_{RT}\sim 1000$ s for the largest turbulent eddies, we ultimately arrive at $\tau_d\sim 3\times 10^4$  s, as observed.

The blackbody, clearly pulsating with the NS spin and with a size compatible with a hot spot, is likely generated on the NS surface. The powerlaw component, pulsating as well as the blackbody component, could be generated by Compton down-scattering in the accretion column shock by the entry of matter through RTI.

\section{Conclusions}

We conducted a comprehensive spectral and timing study of X~Persei using five archival observations obtained with \textit{Chandra} and \textit{XMM-Newton} over a 10 year span. Our key findings are the following:

\begin{itemize}

\item  A two-component continuum, consisting of a hot blackbody ($kT \sim 0.4-1$~keV) and a powerlaw component, is required to describe the spectra. 

The fluxes of the hot blackbody and the powerlaw component are correlated with the NS spin, pointing to the hot spot as their likely origin.\\

\item We find evidence of clump signatures in the Be disk at the NS orbital distance. Time-resolved spectroscopy of over 1200 intervals revealed transient Fe~K$\alpha$ emission that is otherwise diluted in phase-averaged spectra. The prevalence of this line is $\sim 8-9$\%, consistent with the short expected survival times of neutral clumps irradiated by the NS. When detected, the line shows rather consistent properties: $\mathrm{EW}\sim 70$~eV and $F\sim 4\times10^{-4}$~photons~s$^{-1}$~cm$^{-2}$. These detections are related to the presence of clump candidates identified in the light curves. \\

\item We reconstructed the circumstellar disk structure using a viscous decretion disk model. Although the model parameters are degenerate, the best-fit solutions indicate a dense and fairly compact disk, characterized by $\alpha \sim (2.4$–$3.3)$ and $\rho_{0}\sim(6$–$20)\times10^{-10}\,\mathrm{g\,cm^{-3}}$.\\

\item The spin period exhibits a quasi-sinusoidal modulation around $P_{0}\simeq 836~\mathrm{s}$ with $\tau\simeq 3\times10^{4}~\mathrm{s}$ (XMM-2, Chandra-3), consistent with alternating spin-up and spin-down episodes (torque sign changes). The peak rate implied by the modulation is $|\dot P|\sim 10^{-3}~\mathrm{s\,s^{-1}}$, i.e., $|\dot\omega|\sim \mathrm{few}\times10^{-9}~\mathrm{s^{-2}}$. This behavior is compatible with quasi-spherical settling accretion at $R_{\mathrm{A}}\sim10^{10}~\mathrm{cm}$ for $\dot M\sim10^{14}~\mathrm{g\,s^{-1}}$, where the characteristic timescale can be associated with magnetic diffusion near the magnetospheric boundary.

\end{itemize}

\begin{acknowledgements}

GSF, JMTV, JPV and JJRR acknowledge the financial support from the MICIU/AEI/10.13039/501100011033 with funding from the European Union (FEDER). Project (NewAthena24-UA), reference PID2024-155779OB-C33, and from the MCIN with funding from the European Union NextGenerationEU and Generalitat Valenciana in the call Programa de Planes Complementarios de I+D+i (PRTR 2022). Project (Athena-XIFU-UA), reference ASFAE/2022/002. J.J.R.R. acknowledges financial support from the Spanish Ministry of Education, Culture and Sport fellowship PRX23/00270, and also thanks all the staff from SRON for their collaboration and hospitality there. The work of KP was conducted under the state assignment of M.~V.~Lomonosov Moscow State University.
\end{acknowledgements}

\bibliographystyle{aa}
\bibliography{bib} 
\newpage

\newpage

\begin{appendix}
\section{Additional tables.}
\small

\begin{table*}[hb]
  \centering
  \caption{Emission lines detected in the average CCD spectra. }
  \label{tab:appendixA}
  \begin{adjustbox}{max width=0.7\textwidth}
  \begin{tabular}{llcc}
    \toprule
    \textbf{Observation} & \textbf{Most probable ion}\tablefootmark{a} &
    \textbf{Centroid (keV)} &
    \textbf{Flux($\times10^{-5}$ photons s$^{-1}$ cm$^{-2}$)} \\
    \midrule
    \multirow{3}{*}{XMM-2}
& \ion{Al}{xiii}
  & 1.72$\pm 0.01$
  & 80$\pm$10 \\

& \ion{Mg}{xi}
  & 1.59$\pm$0.01
  & 70$\pm$10 \\

& \ion{Fe}{xxii}
  & 1.43$\pm 0.02$
  & 60$\pm$10 \\

& \ion{Si}{xiv}
  & 2.02$\pm 0.02$
  & 60$\pm$10 \\

& \ion{Fe}{xix}
  & 1.27$\pm$0.03
  & 40$\pm$10 \\

    \midrule

    \multirow{2}{*}{Chandra-3}
    & \ion{Ne}{x} 
      & 1.20$\pm 0.01$
      & 9$\pm$3 \\
    
    & \ion{Fe}{xxiv} 
      & 1.12$\pm 0.01$
      & 9$\pm$3 \\
    \bottomrule
  \end{tabular}
  \end{adjustbox}

\end{table*}

\begin{table*}
\centering
\caption{Best-fit parameters from the combined NS spin-resolved spectral analysis. The indices a, b, c, and d correspond to the NS spin phases 0.12--0.37, 0.37--0.62, 0.62--0.87, and 0.87--0.12, respectively.}
\begin{adjustbox}{max width=0.8\textwidth}
\begin{tabular}{lccccccc}
\hline\hline
\\
Spectra & $\chi_{\rm r}^2$ & d.o.f. &
$N_{\rm H}^{\rm LOC}$ & $K_{\rm powerlaw}$ & $\Gamma$ &
$K_{\rm bb2}$ & $kT_{\rm bb}$ \\
&&&
($\times10^{22}$ cm$^{-2}$) &
photons keV$^{-1}$ cm$^{-2}$ s$^{-1}$ ($\times10^{-3}$) & &
$L_{39}/D_{10}^2$ ($\times10^{-3}$) &
(keV) \\
\midrule

Chandra-1a & 1.04 & 750 &
$2.0^{+0.5}_{-0.4}$ &
$48^{+2}_{-3}$ &
$1.9 \pm 0.1$ &
$1.3 \pm 0.4$ &
$1.2 \pm 0.1$ \\\\

Chandra-1b & 0.98 & 830 &
$1.9^{+0.4}_{-0.3}$ &
$55 \pm 3$ &
$1.9 \pm 0.1$ &
$1.9 \pm 0.4$ &
$1.2 \pm 0.1$ \\\\

Chandra-1c & 1.04 & 580 &
$2.3^{+0.9}_{-0.6}$ &
$46 \pm 3$ &
$1.9 \pm 0.2$ &
$1.4^{+0.6}_{-0.4}$ &
$1.1^{+0.1}_{-0.2}$ \\\\

Chandra-1d & 1.06 & 580 &
$4^{+4}_{-2}$ &
$38^{+2}_{-5}$ &
$1.6^{+0.2}_{-0.1}$ &
$0.5^{+0.5}_{-0.3}$ &
$0.9^{+0.5}_{-0.3}$ \\\\

\midrule

XMM-1a & 0.90 & 1340 &
$34^{+8}_{-6}$ &
$26 \pm 2$ &
$0.89^{+0.06}_{-0.05}$ &
$5.0^{+0.4}_{-0.0}$ &
$0.95 \pm 0.04$ \\\\

XMM-1b & 0.83 & 1300 &
$50^{+16}_{-12}$ &
$43  \pm 3$ &
$1.03^{+0.07}_{-0.06}$ &
$6.3 \pm 0.5$ &
$1.04 \pm 0.04$ \\\\

XMM-1c & 0.7 & 1100 &
$1.0 \pm 0.1$ &
$45^{+3}_{-2}$ &
$1.32^{+0.04}_{-0.12}$ &
$1.8^{+0.4}_{-0.3}$ &
$1.3 \pm 0.1$ \\\\

XMM-1d & 0.7 &1030 &
$1.2 \pm 0.3$ &
$33^{+2}_{-4}$ &
$1.3^{+0.2}_{-0.1}$ &
$1.3^{+0.6}_{-0.3}$ &
$1.3^{+0.2}_{-0.1}$ \\\\

\midrule

Chandra-2a & 0.9 & 450 &
$0.8^{+0.4}_{-0.3}$ &
$0.63^{+0.11}_{-0.17}$ &
$0.67 \pm 0.19$ &
$<0.084$ &
$<3$ \\\\

Chandra-2b & 0.9 & 460 &
$1.1^{+0.5}_{-0.4}$ &
$0.80^{+0.12}_{-0.15}$ &
$0.74^{+0.20}_{-0.19}$ &
$0.06^{+0.09}_{-0.06}$ &
$2.4^{+0.6}_{-2.4}$ \\\\

Chandra-2c & 1.1 & 540 &
$1.1^{+0.4}_{-0.3}$ &
$0.7 \pm 0.1$ &
$0.6 \pm 0.1$ &
$<1$ &
$<3$ \\\\

Chandra-2d & 1.1 & 530 &
$0.9^{+0.7}_{-0.4}$ &
$0.4 \pm 0.1$ &
$0.6 \pm 0.2$ &
$<0.008$ &
$<3$ \\\\

\midrule

XMM-2a & 1 & 1000 &
$1.21^{+0.06}_{-0.05}$ &
$35.4 \pm 0.6$ &
$1.18 \pm 0.01$ &
$1.6 \pm 0.1$ &
$1.16 \pm 0.03$ \\\\

XMM-2b & 1.1 & 1200 &
$1.02^{+0.05}_{-0.06}$ &
$51 \pm 1$ &
$1.27^{+0.03}_{-0.04}$ &
$2.1 \pm 0.2$ &
$1.28^{+0.04}_{-0.05}$ \\\\

XMM-2c & 0.9 & 1200 &
$40^{+8}_{-6}$ &
$42 \pm 1$ &
$1.10 \pm 0.03$ &
$3.7 \pm 0.2$ &
$0.96 \pm 0.03$ \\\\

XMM-2d & 1.1 & 1000 &
$22^{+3}_{-2}$ &
$21.1^{+0.6}_{-0.5}$ &
$1.13 \pm 0.02$ &
$1.41 \pm 0.05$ &
$0.73 \pm 0.02$ \\\\

\midrule

Chandra-3a & 0.99 & 1000 &
$70 \pm 20$ &
$25 \pm 1$ &
$1.03 \pm 0.07$ &
$3.7 \pm 0.3$ &
$0.96 \pm 0.02 $ \\\\

Chandra-3b & 1.13 & 1200 &
$62^{+14}_{-13}$ &
$31.3 \pm 1.5$ &
$0.96^{+0.06}_{-0.05}$ &
$4.6 \pm 0.3$ &
$0.96 \pm0.02 $ \\\\

Chandra-3c & 1.13 & 1200 &
$63^{+14}_{-13}$ &
$31.4 \pm 1.5$ &
$0.96^{+0.06}_{-0.05}$ &
$4.6 \pm 0.3$ &
$0.96 \pm 0.02$ \\\\

Chandra-3d & 1.1 &1000 &
$100^{+0}_{-50}$ &
$22 \pm 1$ &
$1.1 \pm 0.1$ &
$2.1 \pm 0.3$ &
$0.99 \pm 0.03$ \\\\

\hline
\end{tabular}
\end{adjustbox}
\label{tab:spin_resolved_params}
\end{table*}

\end{appendix}

\end{document}